\UseRawInputEncoding
\documentclass[prl,aps,twocolumn,superscriptaddress]{revtex4}

\usepackage[tbtags]{amsmath}
\usepackage{amssymb}
\usepackage{amsfonts}
\usepackage{bmpsize}
\usepackage{mathtools}
\usepackage[pdfstartview=FitH]{hyperref}
\hypersetup{colorlinks,linkcolor=blue,filecolor=blue,urlcolor=blue,citecolor=blue,}
\usepackage{multirow}
\usepackage{amsmath,amsfonts,amssymb,times,natbib}

\usepackage[standard]{ntheorem}

\newcommand{\Tr}{\mathrm{Tr}}

\usepackage{physics}

\usepackage{url}
\usepackage{latexsym}
\usepackage{lhelp}

\begin{document}

\title{Entanglement Structure Detection via Machine Learning}

\author{Changbo Chen}
 \affiliation{Chongqing Key Laboratory of Automated Reasoning and Cognition, Chongqing Institute of Green and Intelligent Technology, Chinese Academy of Sciences, Chongqing 400714, People's Republic of China}
\author{Changliang Ren}\thanks{Corresponding author: renchangliang@hunnu.edu.cn}
\affiliation{Key Laboratory of Low-Dimensional Quantum Structures and Quantum Control of Ministry of Education, Key Laboratory for Matter Microstructure and Function of Hunan Province, Department of Physics and Synergetic Innovation Center for Quantum Effects and Applications, Hunan Normal University, Changsha 410081, China}\affiliation{Center for Nanofabrication and System Integration, Chongqing Institute of Green and Intelligent Technology, Chinese Academy of Sciences, People's Republic of China}
\author{Hongqing Lin}\affiliation{Center for Nanofabrication and System Integration, Chongqing Institute of Green and Intelligent Technology, Chinese Academy of Sciences, People's Republic of China}
\author{He Lu}\affiliation{School of Physics, Shandong University, Jinan 250100, China}

%
%

\date{\today}
\begin{abstract}
  Detecting the entanglement structure, such as intactness and depth, of an $n$-qubit state is
  important for understanding the imperfectness of the state preparation in experiments.
  However, identifying such structure usually requires an exponential number of local measurements.
  In this letter, we propose an efficient machine learning based approach for predicting
  the entanglement intactness and depth simultaneously. The generalization ability of this classifier has been convincingly proved,
  as it can precisely distinguish the whole range of pure generalized GHZ states which never exist in the training process. In particular, the learned classifier can discover the entanglement intactness and
  depth bounds for the noised GHZ state, for which the exact bounds are only partially known.
\end{abstract}


\maketitle

Entanglement~\cite{PhysRev.47.777, RevModPhys.81.865}, a pivotal characteristic of quantum theory,
is an important resource for many applications, such as quantum cryptography~\cite{PhysRevLett.67.661},
quantum communication~\cite{RevModPhys.82.665} and quantum computing~\cite{nielsen_chuang_2010,Galindo}. Genuinely multipartite entangled pure state is usually an ideal resource for various quantum information schemes. However, a very important and more realistic situation is that, due to inevitable coupling with the environment,
it is a formidable challenge to produce genuinely multipartite entangled states~\cite{PhysRevX.8.021072}. As a result, the possibility to obtain fewer-body entanglements in the process of preparing $n$-partite entanglement states is high. To characterize such real entanglement structure, it is essential to determine the \emph{entanglement depth}~\cite{PhysRevLett.86.4431} and the \emph{entanglement intactness}~\cite{PhysRevX.8.021072} of these states.

Both entanglement intactness and depth reveal important information on the entanglement structure as well as provide a way for quantifying the entanglement in a multipartite state \cite{PhysRevX.8.021072,Guhne2005}. There have been several previous research focusing on detecting the intactness (or separability)~\cite{PhysRevA.61.042314} and depth (or producibility)~\cite{PhysRevLett.114.190401} of some special multipartite entangled states, such as the noised GHZ state \cite{guhne2008entanglement,Zhou1}. However, the structure of entanglement can be extremely complicated as the number of particles $N$ increases~\cite{PhysRevX.8.021072,Zhou}. The existing research results show that the task of quantifying the entanglement in a multipartite state is not trivial, even for special multipartite entangled states \cite{PhysRevLett.114.190401,PhysRevX.8.021072,PhysRevA.61.042314}. There are at least two challenges for traditional methods: (i) \emph{The current method can not directly and efficiently distinguish different entanglement structure due to the overlap of bounds for different intactness and depth} \cite{PhysRevX.8.021072}. (ii) \emph{These classification methods are state-dependent. In general, it is not known whether one can, simultaneously, classify both entanglement intactness and depth, based on the same set of measurements or observables}~\cite{PhysRevLett.114.190401,PhysRevX.8.021072,PhysRevA.61.042314}.

Contrast to the analytic or numeric approaches, which are model-driven, machine learning (ML) is a data-driven approach
which makes predictions through learning from a large amount of existing data.
Recently, machine learning (ML) has found successful applications on studying many problems in physical sciences~\cite{Carleo2019}.
In particular, there have been some works on studying quantum theory by ML, such as representing quantum states~\cite{Deng2017}, entanglement detection~\cite{Ma2018}, steerability detection~\cite{RenChen2018}, Bell non-locality detection~\cite{PhysRevLett.122.200401, PhysRevLett.120.240402}, etc.

These works motivate us to apply ML for detecting quantum structure. In this letter, through the approach of supervised machine learning, we successfully built efficient classifiers for predicting the multipartite entanglement structure of states composed by random subsystems in very high accuracy. This model has very high prediction and generalization ability. Three important test results strongly support such declaration. Firstly, this model can be used to determine the whole range of generalized $n$-qubit GHZ states with accuracy $\geq 0.999$, which is considered unlikely using device-independent witness for entanglement depth when $n$ is odd \cite{PhysRevA.61.042314}. Secondly, we exhibit that the same data used for building the model for random states can also be used to learn models to predict the entanglement structure of the noised GHZ states, which have no exact known bounds for certain values of intactness and depth. Thirdly, such entanglement structure discriminator generated by theoretical simulation data is directly used to detect the experimental raw data given by \cite{PhysRevX.8.021072}, and the results show that the entanglement structure of these states can be well distinguished. In particular, the intactness of all states are correctly predicted. It is different from various previous work, where the model trained by theoretical-generated data can not predict the obtained experimental data well \cite{PhysRevLett.123.190401,PhysRevLett.120.240501}. As a result, one has to modify the machine learning model. Especially, it should be emphasized that, during the whole classifier-built process, we have never adjusted the parameters intentionally to make the model close to the prediction answer based on experimental data \cite{PhysRevX.8.021072}. This is undoubtedly necessary and valuable to show the validity of predicting the entanglement structure of unknown multipartite entangled states by machine learning.

{\emph{Intactness and Depth.-}} In general, a multipartite system may be a product of several subsystems. If one of the largest subsystems cannot be further decomposed
as products of smaller systems, the number of parties in the largest subsystem is called the entanglement depth. If none of the subsystems can be further decomposed, the number of subsystems is called the entanglement intactness. Thus the depth and intactness of a genuine $n$-partite state are respectively $n$ and $1$ while for a fully separable $N$-partite state, the depth and intactness are respectively $1$ and $n$.

Nevertheless, it is necessary to briefly review them precisely \cite{PhysRevLett.114.190401}. An $n$-partite pure state $\ket{\Phi}$ is said to be $k$-producible, $1\leq k\leq n$, if there exists a decomposition $\ket{\Phi}=\otimes_{i=1}^m \ket{\phi_i}$, where each $\ket{\phi_i}$ is at most $k$-partite. A mixed state $\rho$ is $k$-producible, if it can be written as a convex
combination of $k$-producible pure states. It means that a quantum state is of entanglement depth $k$, if it is $k$-producible, but not $(k-1)$-producible. Similarly, an $n$-partite pure state $\ket{\Phi}$ is said to be $m$-separable if there exists a decomposition $\ket{\Phi}=\otimes_{i=1}^m \ket{\phi_i}$. A mixed state $\rho$ is $m$-separable, if it can be written as a convex combination of $m$-separable pure states. A quantum state is said to be of entanglement intactness $m$, if it is $m$-separable, but not $m+1$-separable.



{\emph{The base model for a class of randomly constructed states}.-} To explore the entanglement structure detection via machine learning, a large amount of labeled data for training is necessary. Two tricky problems need to be resolved in this process. Firstly, it is extremely costly and difficult to obtain these data experimentally especially for multipartite entangled states \cite{PhysRevX.8.021072}. Here, we use simulated data instead. Secondly, if we arbitrarily choose random states, there is no efficient way to detect their entanglement structure, which makes generating a dataset of even moderate size impossible. To overcome this difficulty, we adopt a reverse engineering approach. The $n$-qubit Greenberger-Horne-Zeilinger (GHZ) states,  defined as $\ket{GHZ}=\frac{1}{\sqrt{2}}\left(\ket{0}^{\otimes n}+\ket{1}^{\otimes n}\right)$, are used as seeds to generate the training data.
The genuinely entangled state of such class can be distinguished, where the operator ${\cal W}_{G}=\frac{1}{2}{\mathbb I}-\ket{GHZ}\bra{GHZ}$ defines a witness for determining if a $n$-partite mixed state $\rho$ belongs to the GHZ class~\cite{PhysRevLett.92.087902}. Namely if $\Tr[{\cal W}_G\rho]<0$, then $\rho$ belongs to the GHZ class and is thus genuinely entangled.

Since for any $m$, we already have criterion determining if a noised $m$-qubit GHZ state is genuinely entangled. Then for any $n$, we can build a state with known intactness and depth
with these $m$-qubit states, where $m=1,\ldots,n$. When $n$ increases, the number of partitions also increases. To reflect this complexity, the size of the dataset for learning the model also increases.

To summarize, we start by creating some seed states, which are genuinely entangled $n$-partite states for $n\geq 2$.
With these seed states as basic building blocks, we construct $n$-partite states by exhausting all possible combinations of intactness and depth.
These states serve as our training states. Note that since the entanglement intactness and depth of each training state are already known by construction, we can naturally assign a label to each state by encoding both the entanglement intactness and depth information. By creating a large number of such random states, we obtain a large dataset, from which we learn a neural network classifier.




For $n=1,\ldots,12$, we define some seed states as below
\begin{equation}
\rho_{n}=(1-\alpha-\beta)\tau_{n}+\alpha \eta_n+\beta {\mathbb I}_n/2^n,
\end{equation}
where $\tau_n=\ket{G}\bra{G}$, $\ket{G}=\frac{1}{\sqrt{2}}\left(\ket{0}^{\otimes n}+\ket{1}^{\otimes n}\right)$,
$\eta_n=\frac{1}{2}\left( \ket{0}^{\otimes n}\bra{0}^{\otimes n} +\ket{1}^{\otimes n}\bra{1}^{\otimes n} \right)$,
${\mathbb I}_n$ is the $2^n\times 2^n$ identity matrix,
and $\alpha,\beta\in [0,1]$ satisfy that $\Tr[{\cal W}_G\rho_n]<0$, that is $\frac{1}{2}(-1+\alpha)+\frac{2^n-1}{2^n}\beta<0$.
In particular, when $n=1$, an equivalent form of $\rho_1$ is $\rho_{1}=(1-\alpha)\tau_{1}+\alpha {\mathbb I}_1/2$.

With such seed states, we prepare our training states recursively according to the following partition:
\begin{equation}
  {\frak P}(n) = \{[n]\}\cup_{i=1}^{n-1}\{{\rm cat}(a, b)\mid a\in {\frak P}(i), b\in {\frak P}_{n-i}\},
\end{equation}
where ${\rm cat}(a, b)$ is a concatenation of two lists and duplicate elements are removed from the set ${\frak P}(n)$.
The number of elements in ${\frak P}(n)$ is $2^{n-1}$.
For each element $L$ of ${\frak P}(n)$, the training state is defined as $\xi=\otimes_{\ell \in L}\rho_{\ell}$.
The intactness $m$ and depth $d$ of $\xi$ are respectively the number of elements and the maximum element in $L$.
For a given $n$, we sort all possible values of $(m,d)$ in an ascending lexicographic order.
For a given $\xi$, the position of its $(m,d)$ in the new ordering is set as its label.
For each $n=4,\ldots,12$, the number of total labels is
\begin{equation}
  \label{eq:class}
{\frak n}_{\ell} := (n^2+3n)/2-1-\sum_{i=1}^n\lceil n/i \rceil.
\end{equation}
In such experimentation, for each $L$, we create $15000$ such states $\xi$, for random values of $\alpha$ and $\beta$,
$2/3$ of which are reserved for the training dataset, $1/6$ for the validation dataset,
and the rest for the testing dataset.

Instead of choosing the whole information of these training states as feature, partial information are used.
For each state $\rho$, we construct its feature vector $(\langle \hat{O}_1\rangle, \ldots, \langle\hat{O}_s\rangle)$ by computing $\langle \hat{O}_i\rangle = \Tr[\hat{O}_i\rho]$,
for a set of well-chosen operators $\hat{O}_1,\ldots,\hat{O}_{s}$.
In this work, the following operators~\cite{PhysRevX.8.021072} are chosen:
\begin{equation}\label{oper}
\begin{array}{rcl}
  M_{z} &=& (\ket{0}\bra{0})^{\otimes n}+(\ket{1}\bra{1})^{\otimes n}\\
  M_{x} &=& \sigma_{x}^{\otimes n}\\
  A_{z} &=& A_{+}^{\otimes n},\\
  A_{x} &=& (\frac{A_{+}+A_{-}}{2})^{\otimes n}\\
\end{array}
\end{equation}
where $A_{+}=\cos{(\frac{n+1}{2n}\phi_{n})}\sigma_x+\sin{(\frac{n+1}{2n}\phi_{n})}\sigma_y$ and $A_{-}=\cos{(\frac{-(n-1)}{2n}\phi_{n})}\sigma_x+\sin{(\frac{-(n-1)}{2n}\phi_{n})}\sigma_y$.
Here, $\phi_{2}=\frac{\pi}{2}$, $\phi_{3}=1.231$, $\phi_{4}=1.0155$, $\phi_{5}=0.866$, $\phi_{6}=0.7559$, $\phi_{7}=0.6713$~\cite{PhysRevLett.114.190401}, $\phi_{8}=0.6$~\cite{PhysRevX.8.021072}, and $\phi_{n}=\frac{2\pi}{n}$ for $n>9$~\cite{PhysRevLett.114.190401}. In this way, we have generated a large amount of data for training and testing an entanglement structure discriminator.

\begin{figure}[ht]
  \begin{minipage}{0.5\linewidth}
    \centering
    $(a)$\\
    \includegraphics[width=0.65\linewidth]{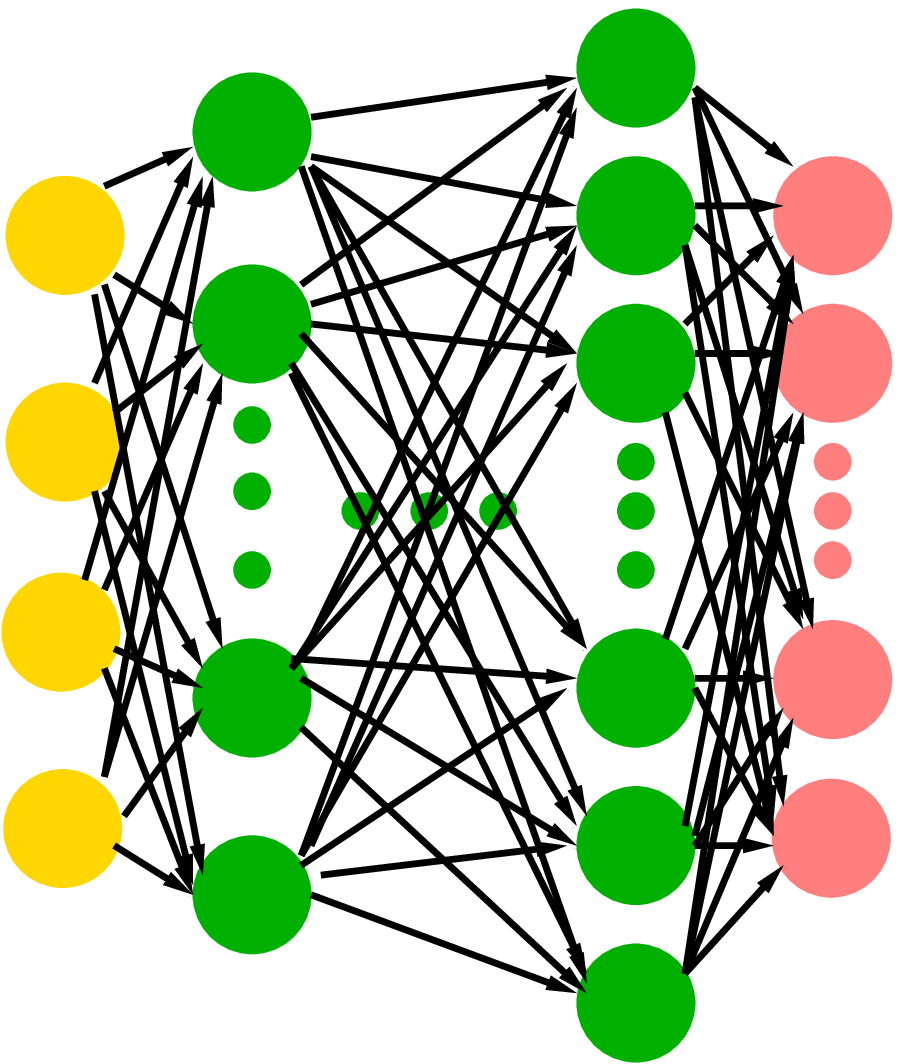}
  \end{minipage}
  \begin{minipage}{0.5\linewidth}
    \centering
    $(b)$\\
    \includegraphics[width=\linewidth]{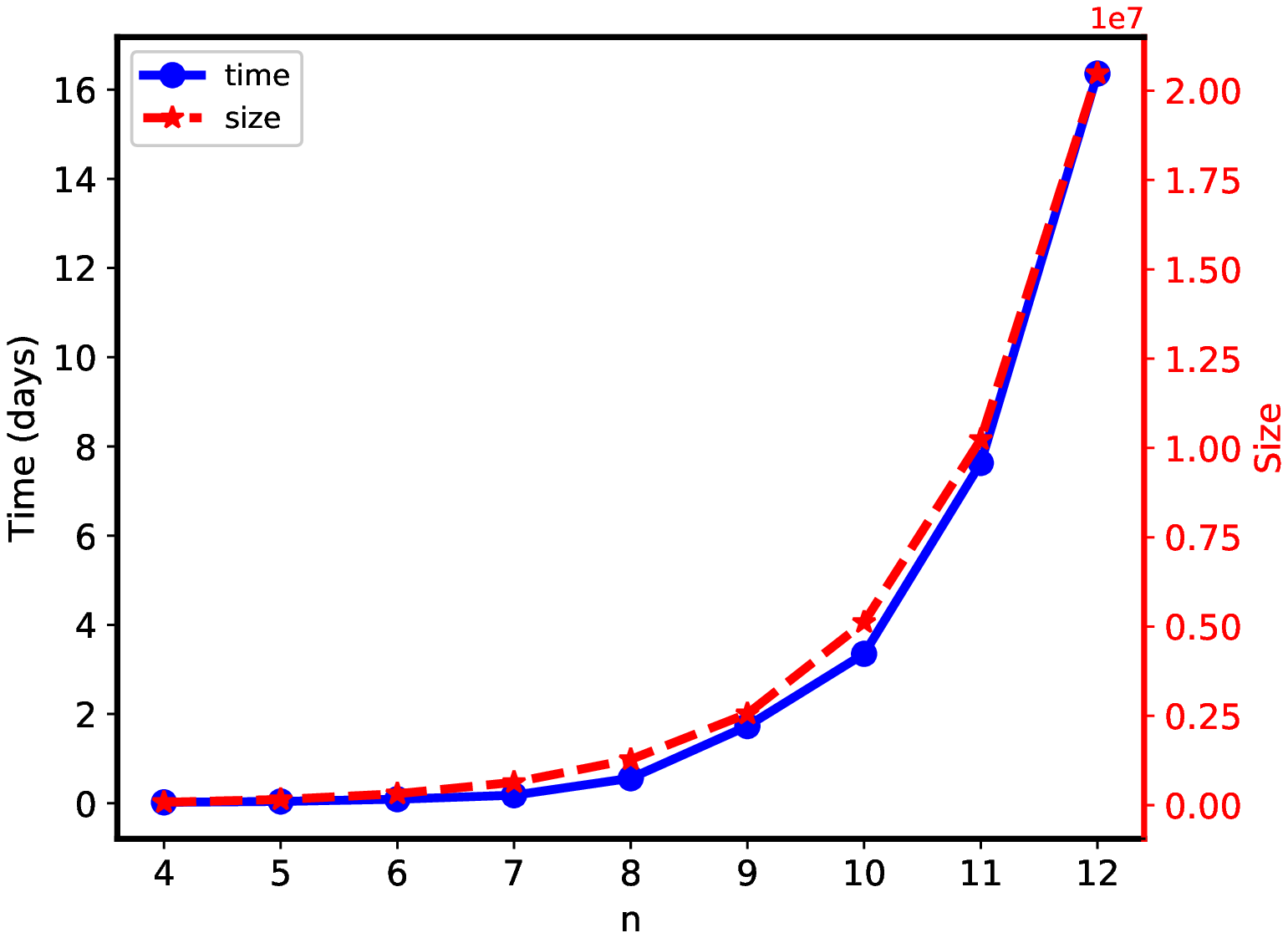}
  \end{minipage}
  \begin{minipage}{0.5\linewidth}
    \centering
    $(c)$\\
    \includegraphics[width=\linewidth]{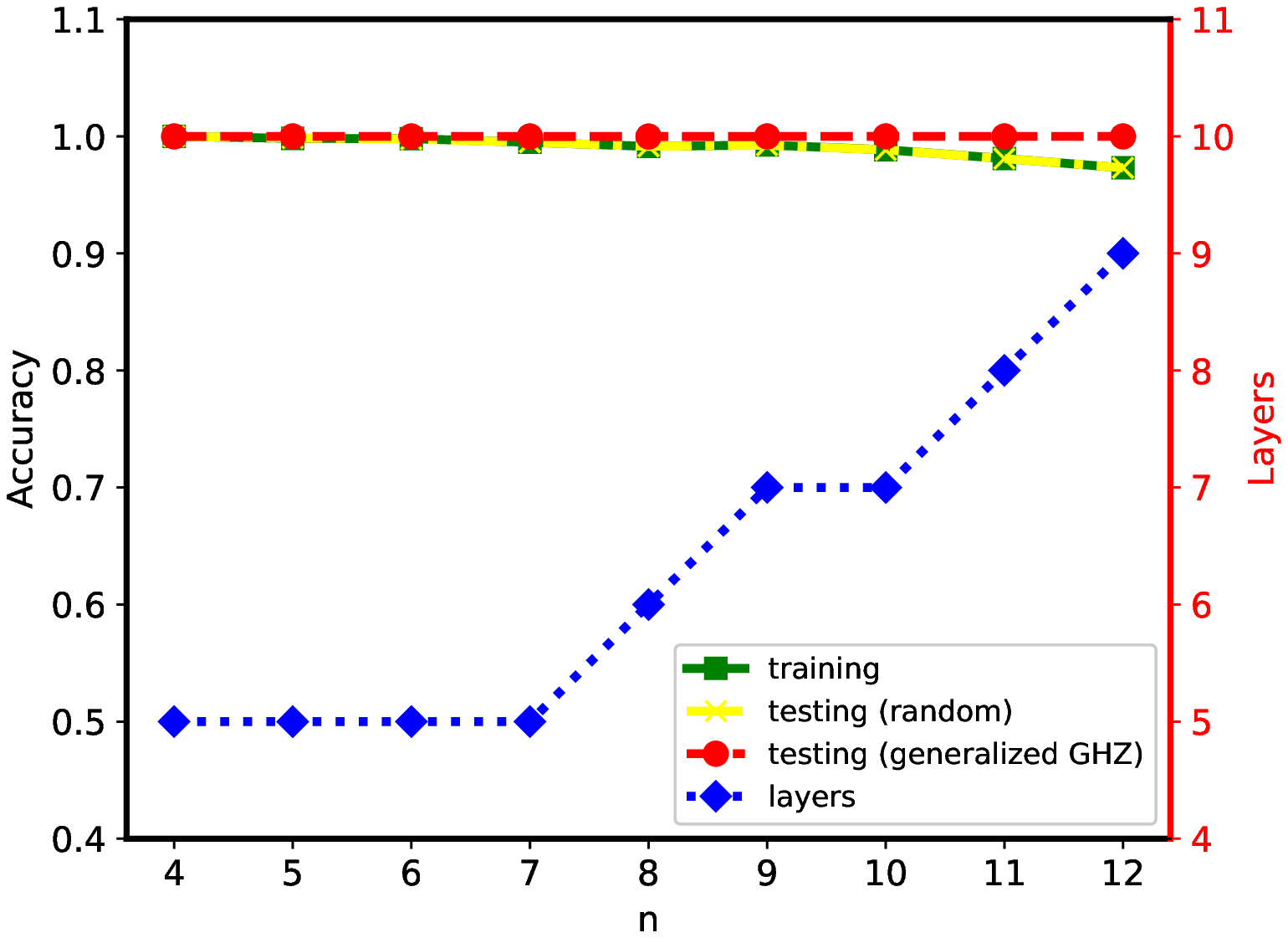}
  \end{minipage}
  \begin{minipage}{0.5\linewidth}
    \centering
    $(d)$\\
    \includegraphics[width=\linewidth]{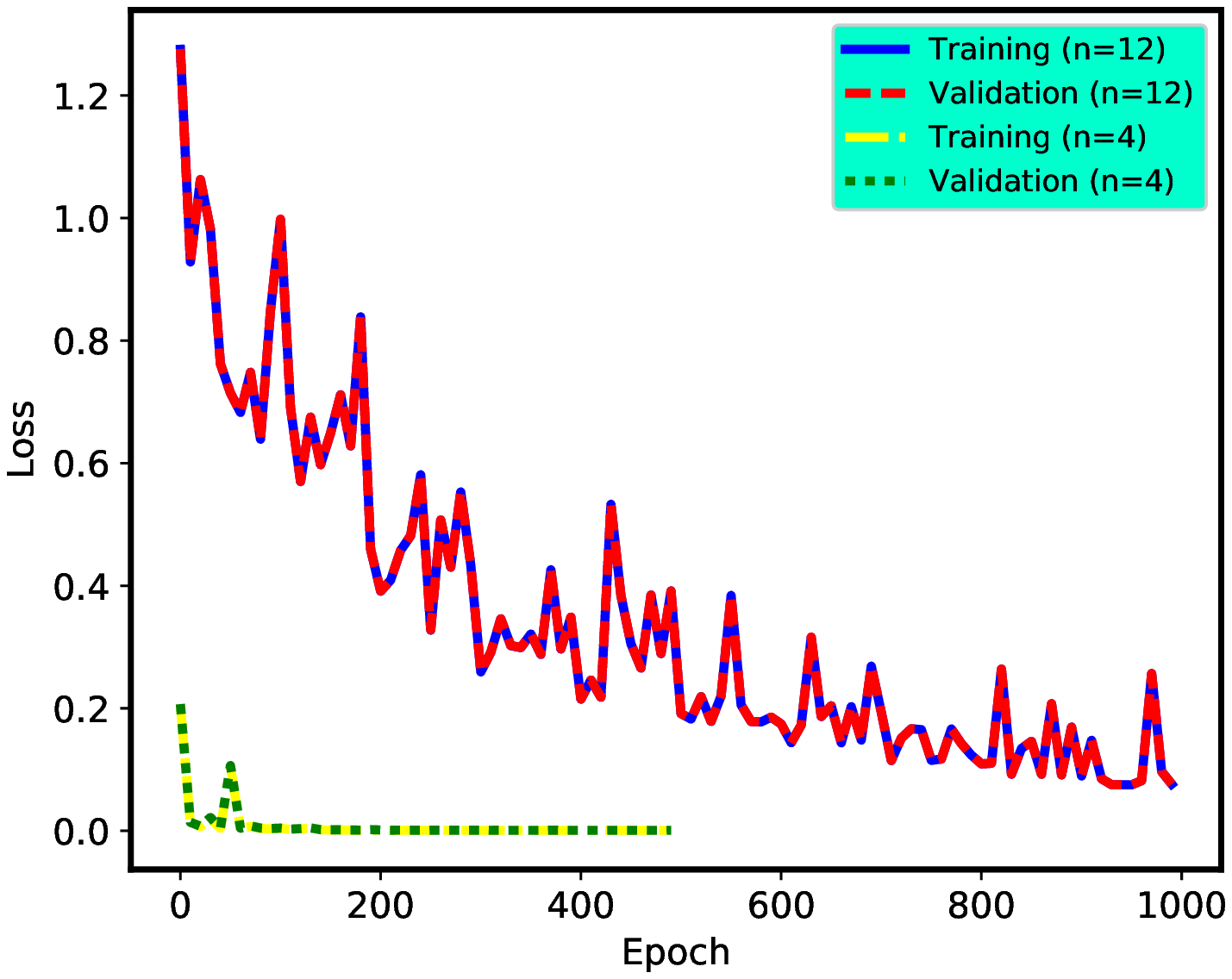}
  \end{minipage}
  \caption{\label{fig:base}(Color online) \small{The experimental results with the base random model.
      Fig. $(a)$ depicts the architecture of the neural networks for the random model.
      Fig. $(b)$ illustrates the training time (blue $\cdot$) and the size (red $\star$) of the  training dataset for each $n=4,\ldots, 12$.}
    Fig. $(c)$ reports the training accuracy (green $\square$), the testing accuracy on the random dataset (yellow $\cross$), the testing accuracy on the dataset of pure GHZ states (red $\cdot$),
  as well as the number of layers of the neural networks (blue $\diamond$). Fig. $(d)$ demonstrates the values of loss functions during training and validation for $n=4$ (two bottom overlapped lines) and $n=12$ (two top overlapped lines).}
\end{figure}

For each $n=4,\ldots,12$, we employ a neural network with one input layer, one softmax output layer,
and several hidden layers, whose active functions are ReLu~\cite{Goodfellow-et-al-2016}, illustrated in Fig.~\ref{fig:base} $(a)$.
The number of total layers is illustrated in Fig.~\ref{fig:base} $(c)$.
As $n$ increases, the neural network also becomes more complex (with more hidden layers and more neurons per layer).
It is interesting to observe from Fig.~\ref{fig:base} $(b)$ that the training time increases linearly with the training dataset size ($2^{n-1}\times 10^4$),
while both are exponential functions of $n$. For $n=12$, the dataset contains more than 20 million examples.
As we can see from Fig.~\ref{fig:base} $(c)$, as $n$ increases, the testing accuracy for the reserved testing dataset (with the same distribution as the training dataset)
drops a little, but still higher than $0.97$.
More importantly, the fact that the training accuracy and testing accuracy are almost identical shows that
the model has quite good generalization ability on unseen data, which can also be justified
by the values of loss functions shown in Fig.~\ref{fig:base} $(d)$.
In Fig.~\ref{fig:base} $(d)$, we only show the values of loss functions for $n=4$ and $n=12$.
But for other values of $n$, we see a similar phenomenon, namely the values of loss functions for training
and validating are almost identical and both converge to a low value. When $n$ increases, we pay more effort to train the model.
For instance, we tried $500$ epoches for $n=4$, but $1000$ epochs for $n=12$.
The number of epoches for $n$, $4<n<12$, is between $500$ and $1000$.
Once a model is successfully learned, predicting the entanglement structure of a state with the model
is instantaneous.

In order to further exhibit the power of such machine learning classifier, we use it to distinguish the pure generalized GHZ states, $\cos(\theta)\ket{0}^{\otimes n}+\sin(\theta)\ket{1}^{\otimes n}$, $0\leq \theta< \frac{\pi}{4}$, which never exist in the training process. And what's more, for $n\geq 3$ and $n$ is odd, such states cannot violate any full correlation Bell inequality \cite{PhysRevA.61.042314} which means that the whole range of such states cannot be completely classified by any full correlation Bell inequality \cite{PhysRevA.61.042314,PhysRevLett.114.190401}.

For each  $n=4,\ldots,12$,
we evenly sample $10001$ values from the interval $[0,\frac{\pi}{4}]$ to generate many pure generalized GHZ states and obtain a testing dataset different from the random one.
With the neural network models trained above, we can predict simultaneously the intactness and depth  of these new states.
It turns out that the testing accuracy for these new datasets are all $\geq 0.999$.
In fact, the neural network models only make wrong predictions for one value of $\theta$, namely $\theta=0$, among $10001$ values of $\theta$,
for all $n=4,\ldots,12$, except for $n=8$, where the classifier makes two wrong predictions including the one for $\theta=0$.
Fig.~\ref{fig:base} $(c)$ illustrates the testing accuracy.
This shows that the trained model has good generalization ability even on a testing dataset with potentially different distributions as the training dataset.

\emph{Finding the intactness and depth bounds for the noised GHZ state.-} One may wonder how the model performs on states in real applications which we do not know their entanglement structures at all, for instance on the parameterized noised GHZ states at particular values of the parameters. Note that unlike the random test dataset, the dataset formed by these states may potentially have different distributions as the training set.
The disagreement between distributions of training dataset and testing dataset is one of the main
challenges in applied machine learning~\cite{jin2020domain}. Such issues widely appear in real applications where labels for testing data are hard or expensive to obtain,
for instance in drug discovery~\cite{Strokes2020}, materials science~\cite{Schmidt2019}, and physical sciences~\cite{Carleo2019}.
Next we demonstrate how to train a new model, named GHZ-model, with the previously constructed random training dataset
and use it to predict the intactness and depth bound for a $n$-qubit noised GHZ state
\begin{equation}
  \label{eq:noisy-GHZ}
  \rho_{ng} = p\ket{GHZ}\bra{GHZ}+(1-p){\mathbb I}_n/2^n, 0\leq p\leq 1.
\end{equation}
That is for each given $k$, $1\leq k\leq n$,
we would like to know the range of values of $p$ such that the intactness
or depth of $\rho_{ng}$ is $k$.
Or equivalently, for each $k$, we would like to find a bound $b_k$
such that $\rho_{ng}$ is $k$-separable (or $k$-producible) if and only if $p\leq b_k$.

For separability, precise analytic bounds are well-known for $k=2$~\cite{guhne2008entanglement}, namely $p\leq\frac{2^{n-1}-1}{2^n-1}$
and $k=n$~\cite{PhysRevA.61.042314}, namely $p\leq \frac{1}{1+2^{n-1}}$.
The paper~\cite{CJX2018} proves an analytic bound for $k\geq \frac{n+1}{2}$, that is $p\leq\frac{1}{1+\frac{2k-n}{n}2^{n-1}}$.
To our best knowledge, no precise analytic separability bounds are known for $2<k<\frac{n+1}{2}$.
There are no precise analytic producibility bounds either for general values of $k$ (See supplementary materials for details \cite{supplementary}).

The lack of analytic bounds for some values of $k$ is the main remarkable feature of this task. This does create a difficulty to evaluate the quality of learned ML model since
the precise intactness and depth are unknown for many values of $p$. We overcome this difficulty by redefining ``correct predictions''.

Recall that for a given $n$, we have trained a multiclassifier ${\cal C}$ which may classify
a $n$-qubit state $\rho$ into a class among $\{0,\ldots,{\frak n}_{\ell}-1\}$.
Normally, if the class of a state is known, say $c$, we say ${\cal C}$ makes a correct prediction
if and only if ${\cal C}(\rho)=c$.
However, for the noised GHZ state $\rho_{ng}$, its class may be unknown for some values of $p$.
Thus, we re-define ``correct prediction'' as follows.
We say ${\cal C}$ makes a correct prediction for $\rho_{ng}$ if the corresponding intactness $k$ of ${\cal C}(\rho_{ng})$:
\begin{itemize}
\item Either agrees exactly with the intactness of $\rho_{ng}$ for $k=1$ and $k\geq \frac{n+1}{2}$;
\item Or falls in the same range $(1, \frac{n+1}{2})$ as the intactness of $\rho_{ng}$.
\end{itemize}
The accuracy of ${\cal C}$ would be defined as the percentage of correct predictions in the above sense
over all predictions.

It is natural to ask how the models learned in the previous section
for a class of randomly constructed states perform on the noised GHZ states.
To evaluate the model, we evenly sample $10001$ values from $[0, 1]$ for $p$
and formulate a testing dataset with $10001$ examples of $\rho_{ng}(p)$,
whose feature vectors are generated according to Eq. (\ref{oper}).

\begin{figure}[ht]
  \begin{minipage}{\linewidth}
  \begin{minipage}[t]{0.49\linewidth}
    $(a)$\\
    \includegraphics[width=\linewidth]{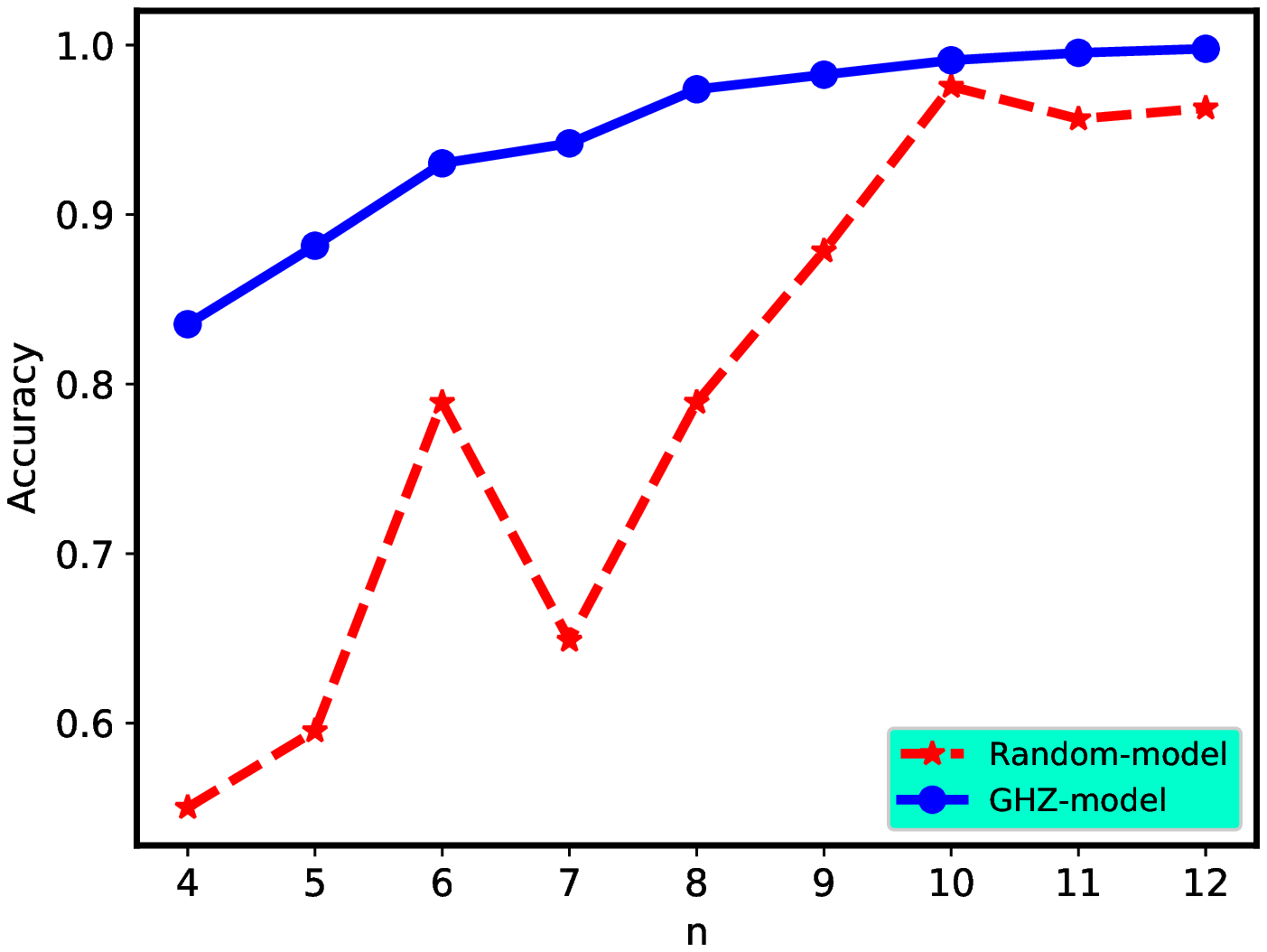}
  \end{minipage}
  \begin{minipage}[t]{0.5\linewidth}
    $(b)$\\\vspace{2ex}
    \includegraphics[width=0.95\linewidth]{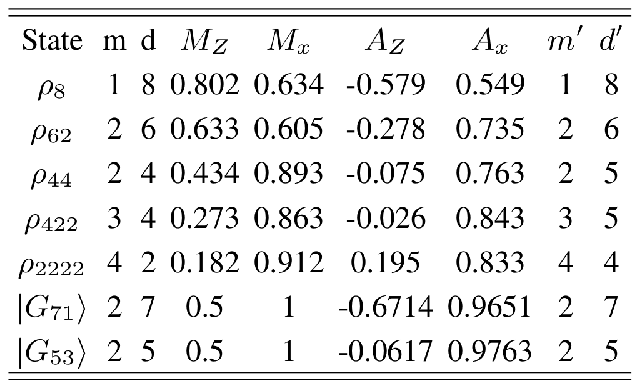}
  \end{minipage}
  \end{minipage}
  \begin{minipage}[t]{0.5\linewidth}
    \centering
    $(c)$\\
    \includegraphics[width=\linewidth]{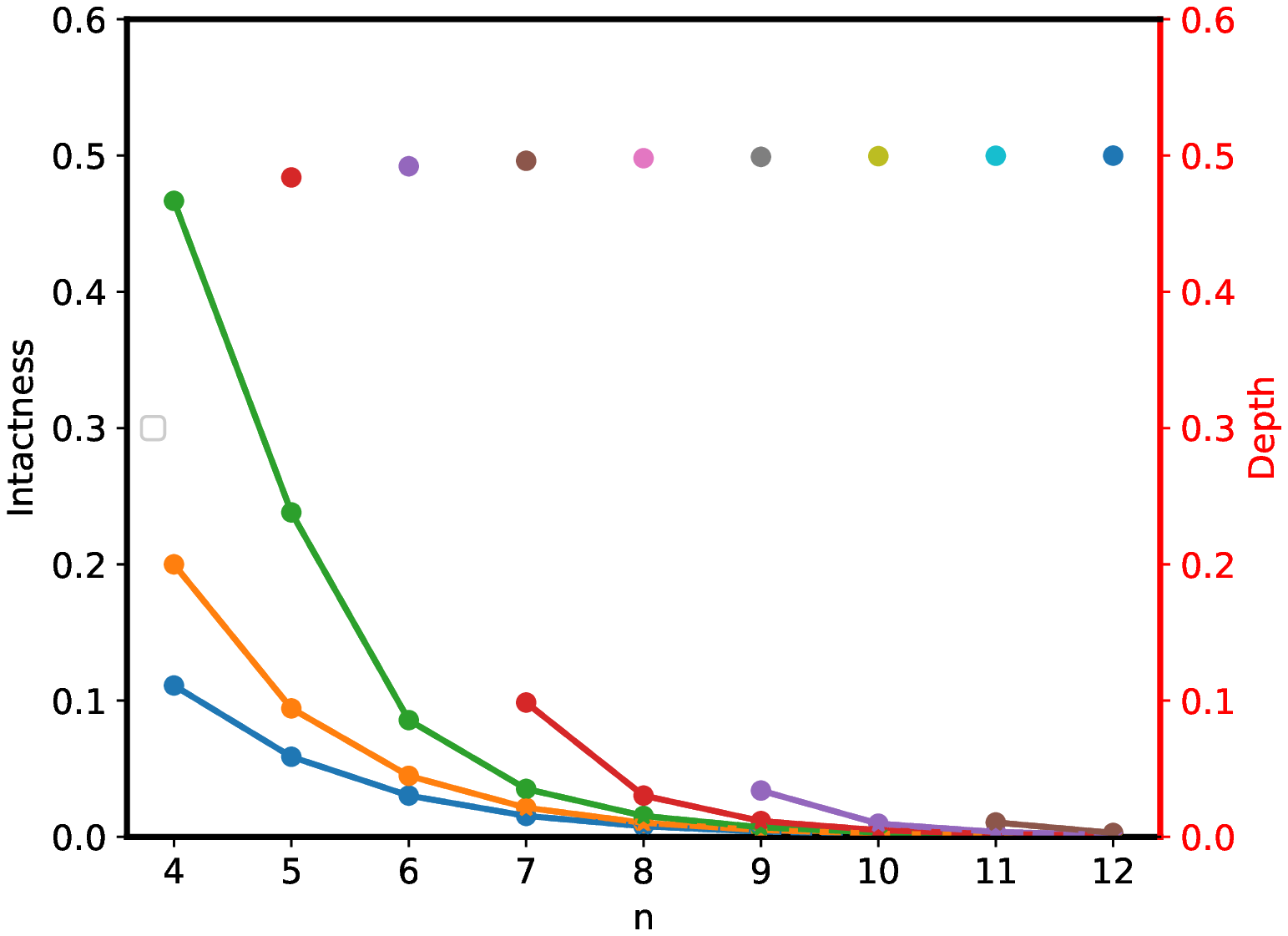}
  \end{minipage}
  \begin{minipage}[t]{0.5\linewidth}
    \centering
    $(d)$\\
    \includegraphics[width=\linewidth]{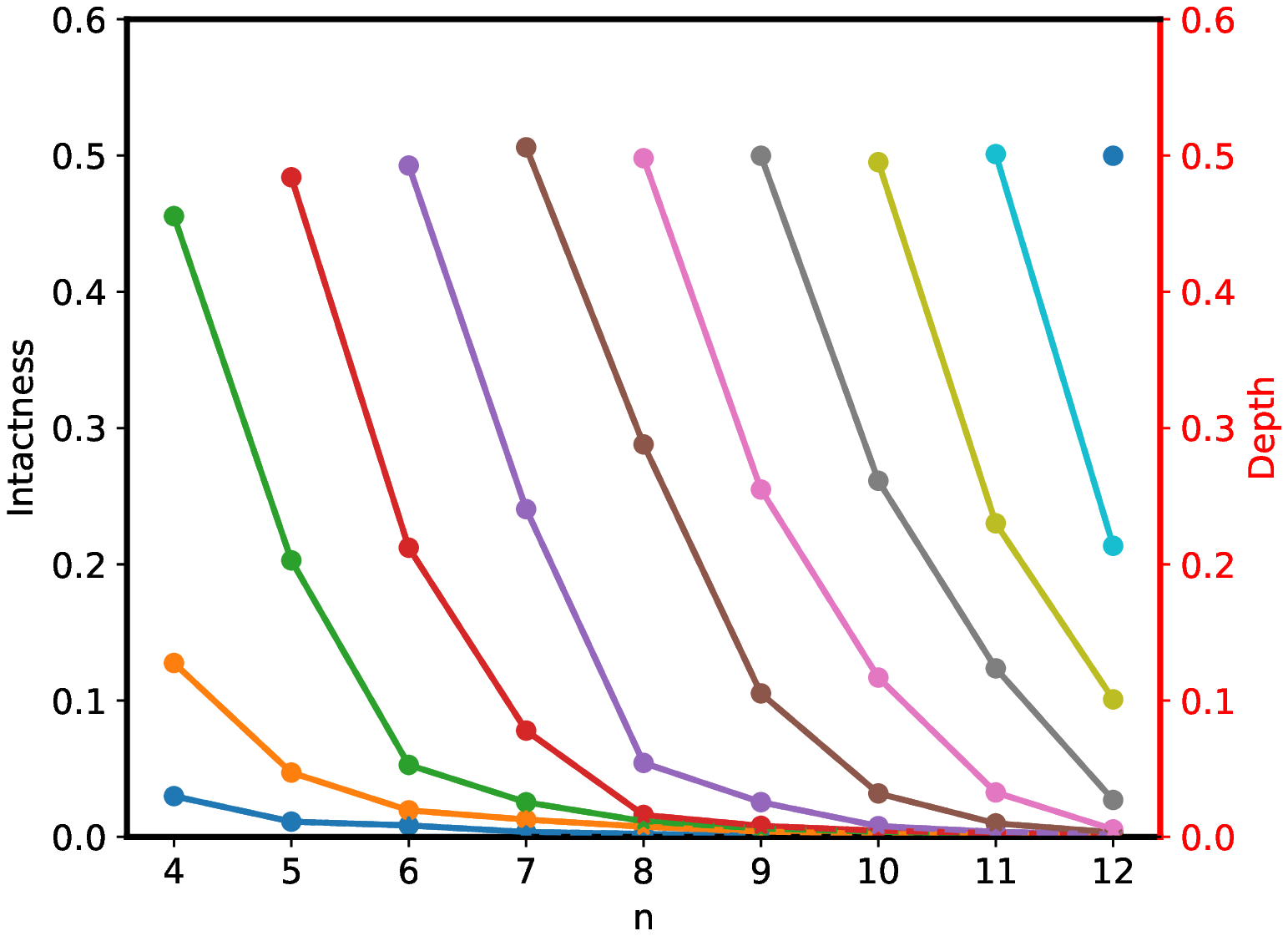}
  \end{minipage}
  \caption{\label{fig:ghz}(Color online) \small{The experimental results on the noised $n$-qubit  GHZ states, $4\leq n\leq 12$.
      Fig. $(a)$ compares the testing accuracies on the noised GHZ states
      between the random-model (red $\star$) and the GHZ-model (blue $\cdot$).
      Fig. $(b)$ (the table) reports testing results on experimental data.
    Fig. $(c)$ illustrates the known analytic intactness and depth bounds,
    where from left to right, the $k$-th line (missing exact bounds are not shown and thus the lines are discontinuous) represents the  exact $k$-depth and $n-k+1$-intactness bounds. Fig. $(d)$ demonstrates the learned bounds by ML, where from left to right, the $k$-th line
    (the $11$-th line degenerates into a point) represents the predicted $k$-depth and $n-k+1$-intactness bounds.  }}
\end{figure}


The red $*$ in Fig.~\ref{fig:ghz} $(a)$ shows the accuracy for the random model.
From this figure, we can see that directly using the model does not work very well.
In other words, the model in the previous section has high generalization error on the testing dataset
for the noised GHZ states. This indicates that the distribution of the dataset for the noised GHZ states may be quite different from that for the constructed random states.

In machine learning, there are several ways to address this issue.
One typical way is to augment the training dataset with data drawn from the same distribution
as the  testing dataset in order to make the two distributions as close as possible.
However, in our case, this approach is not a good choice since it may be difficult to know
the precise label of a noised GHZ state. Note that the models in the previous section is selected through a validation set with the same distribution
as the training set. One may instead create a validation set with similar distribution as the testing set
and use it to select the model (with highest validation accuracy) among epoches.
Moreover, a high error on the testing dataset for the noised GHZ states may indicate that
the trained model may be over-parametrized and it is always a good idea in machine learning
to choose simpler model to reduce the generalization error.
The new neural network model has only one hidden layer with number of neurons as $2^{n+1}$. To further reduce the generalization error, we introduce a regularization term
with the weight decay as $0.1^{n-1}$.

The blue $\boldsymbol{\cdot}$ in Fig.~\ref{fig:ghz} $(a)$ reports the accuracy using this approach.
The new model has quite high accuracy on the testing dataset for the noised GHZ states.
Next we use the model to predict the bound.
Recall that we have generated $10001$ different states $\rho_{ng}(p)$
with the values of $p$ evenly taken from $[0,1]$.
Let $i_p$ (resp. $d_p$) be the predicted intactness (resp. depth) of $\rho_{ng}(p)$.
Then for each $k=1,\ldots,n$, the learned $i$-intactness (resp. $d$-depth) bound
is defined as the largest $p$ such that $i=i_p$ (resp. $d=d_p$).
The learned intactness and depth bounds are given in Fig.~\ref{fig:ghz} $(d)$.
As a contrast, Fig.~\ref{fig:ghz} $(c)$ provides exact bounds that we know so far.



Obviously, the bounds predicted by the ML approach coincide with the known exact bounds quite well.
More importantly, it provides the missing bounds (See supplementary materials for details \cite{supplementary}).
Notice that when $n=10,11,12$, the bounds for depth $1$ is missing.
Moreover, for $n=11,12$, the bounds for depth $2$ is slightly different from
the bound for intactness $n-2$.
For the rest, the depth and intactness bounds exactly agree.
As a final note, for $n=4,\ldots,11$, we always have $p_1\leq p_2$ if and only if $i_{p_1}\geq i_{p_2}$
for intactness (resp. $d_{p_1}\leq d_{p_2}$ for depth).
That is the learned bound is truly a bound.
But for $n=12$, the classifier misclassifies the intactness (resp. depth) of $\rho_{ng}(p), p\in (0.0034, 0.0054)$
and $\rho_{ng}(p), p\in (0.0056, 0.0270)$ from $6$ as $3$ and from $5$ as $4$ (resp. from $7$ as $10$ and $8$ as $9$).

One might be curious on how the model performs on experimental generated states.
To this end, we collect the experimental data reported in~\cite{PhysRevX.8.021072}
into a testing dataset, which consists of $7$ examples.
Fig.~\ref{fig:ghz} $(b)$ reports our testing results with the trained GHZ-model.
In the table, the first column shows the state
together with its structure as index.
We use $m$ and $d$ to denote the true intactness and depth.
The predicted intactness and depth are respectively denoted by $m'$ and $d'$.
The experimentally measured values $M_{z}, M_{x}, A_{z}, A_{x}$ are from~\cite{PhysRevX.8.021072}.
The model predicts the intactness and depth of the experimentally prepared states simultaneously. As we can see from the table, the intactness of all $7$ states have been correctly predicted while
the depth of $3$ states have higher predicted values than their ground truth.

\emph{Conclusion.-} In summary, through the approach of supervised machine learning,
we have successfully built efficient classifiers for predicting
the multipartite entanglement structure of states composed by random
subsystems in very high accuracy.
The same models can also be used to determine if a pure generalized GHZ
state is genuinely multipartite entangled with accuracy $\geq 0.999$.
More interestingly, the same data used for building
the model for random states can also be used to learn
models to predict the entanglement structure of the noised GHZ states,
for which there are no exact known bounds for certain
values of intactness and bounds. Finally, these models were successfully applied to detect
the intactness of $8$-qubit states prepared in experiments.

\section{Acknowledgments}
C. Chen is supported by NSFC (No. 11771421), CAS ``Light of West China'' Program, National Key Research and Development Program (No. 2020YFA0712300), and Chongqing Programs (No. cstc2018jcyj-yszxX0002, No. cstc2019yszx-jcyjX0003, No. cstc2020yszx-jcyjX0005). C. Ren is supported by the Natural Science Foundation of China (No. 12075245), National Key Research and Development Program (No. 2017YFA0305200), and Xiaoxiang Scholars Program of Hunan Normal University.


\begin{thebibliography}{39}
\bibitem{PhysRev.47.777} A. Einstein, B. Podolsky, and N. Rosen, Phys. Rev. 47, 777
(1935).
\bibitem{RevModPhys.81.865} R. Horodecki, P. Horodecki, M. Horodecki, and K. Horodecki, Rev. Mod. Phys. 81, 865 (2009).
\bibitem{PhysRevLett.67.661} A. K. Ekert, Phys. Rev. Lett. 67, 661 (1991).
\bibitem{RevModPhys.82.665}  H. Buhrman, R. Cleve, S. Massar, and R. de Wolf, Rev. Mod. Phys. 82, 665 (2010).
\bibitem{nielsen_chuang_2010}  M. A. Nielsen and I. L. Chuang, Quantum Computation and Quantum Information: 10th Anniversary Edition (Cambridge University Press, 2010).
\bibitem{Galindo} A. Galindo and M. A. Mart\'{\i}n-Delgado, Rev. Mod. Phys. 74, 347 (2002).
\bibitem{PhysRevX.8.021072}  H. Lu, Q. Zhao, Z.-D. Li, X.-F. Yin, X. Yuan, J.-C. Hung, L.- K. Chen, L. Li, N.-L. Liu, C.-Z. Peng et al., Phys. Rev. X 8, 021072 (2018).
\bibitem{PhysRevLett.86.4431} A. S. S{\o}rensen and K. M{\o}lmer, Phys. Rev. Lett. 86, 4431 (2001).
\bibitem{Guhne2005} O. G¨¹hne, G. T\'{o}th, and H. J. Briegel, New J. Phys. 7, 229 (2005).
\bibitem{PhysRevA.61.042314} W. D\"{u}r and J. I. Cirac, Phys. Rev. A 61, 042314 (2000).
\bibitem{PhysRevLett.114.190401} Y.-C. Liang, D. Rosset, J.-D. Bancal, G. P\"{u}tz, T. J. Barnea, and N. Gisin, Phys. Rev. Lett. 114, 190401 (2015).
\bibitem{guhne2008entanglement} O. G\"{u}hne and G. T¨®th, Phys. Rep. 474, 1 (2009).
\bibitem{Zhou1} Y. Zhou, Phys. Rev. A 101, 012301 (2020).
\bibitem{Zhou} Y. Zhou, Q. Zhao, X. Yuan and X. F. Ma,  Npj. Quantum Inform. 5, 83 (2019).
\bibitem{Carleo2019} G. Carleo, I. Cirac, K. Cranmer, L. Daudet, M. Schuld, N. Tishby, L. Vogt-Maranto, and L. Zdeborov¨¢, Rev. Mod. Phys. 91, 045002 (2019).
\bibitem{Deng2017} D. L. Deng, X. Li, and S. Das Sarma, Physical Review X 7, 021021 (2017).
\bibitem{Ma2018} Y. Ma and M. Yung, Npj. Quantum Inform. 4, 34 (2018).
\bibitem{RenChen2018} C. Ren and C. Chen, Phys. Rev. A 100, 022314 (2018).
\bibitem{PhysRevLett.122.200401} A. Canabarro, S. Brito, and R. Chaves, Phys. Rev. Lett. 122, 200401 (2019).
\bibitem{PhysRevLett.120.240402} D.-L. Deng, Phys. Rev. Lett. 120, 240402 (2018).
\bibitem{PhysRevLett.123.190401} M. Yang, C.-L. Ren, Y.-c. Ma, Y. Xiao, X.-J. Ye, L.-L. Song, J.-S. Xu, M.-H. Yung, C.-F. Li, and G.-C. Guo, Phys. Rev. Lett. 123, 190401 (2019).
\bibitem{PhysRevLett.120.240501} J. Gao, L.-F. Qiao, Z.-Q. Jiao, Y.-C. Ma, C.-Q. Hu, R.-J. Ren, A.-L. Yang, H. Tang, M.-H. Yung, and X.-M. Jin, Phys. Rev. Lett. 120, 240501 (2018).
\bibitem{PhysRevLett.92.087902} M. Bourennane, M. Eibl, C. Kurtsiefer, S. Gaertner, H. Weinfurter, O. G\"{u}hne, P. Hyllus, D. Bruss, M. Lewenstein, and A. Sanpera, Phys. Rev. Lett. 92, 087902 (2004).
\bibitem{Goodfellow-et-al-2016} I. Goodfellow, Y. Bengio, and A. Courville, Deep Learning (MIT Press, 2016).
\bibitem{jin2020domain} W. Jin, R. Barzilay, and T. Jaakkola, Arxiv:2006.03908.
\bibitem{Strokes2020} J. Stokes et al., Cell 180, 688 (2020).
\bibitem{Schmidt2019} J. Schmidt, M. Marques, S. Botti, et al., Npj. Comput. Mater. 5, 83 (2019).
\bibitem{CJX2018} X.-Y. Chen, L.-Z. Jiang, and Z.-A. Xu, Int. J. Quantum Inf. 16, 1850037 (2018).
\bibitem{supplementary} Supplementary Materials.
\end{thebibliography}

\end{document}